\documentclass[runningheads]{llncs}
\usepackage[T1]{fontenc}
\usepackage{graphicx}
\usepackage{booktabs}

\usepackage{cite}
\usepackage{amsmath,amssymb,amsfonts}
\usepackage{algorithmic}
\usepackage{textcomp}
\usepackage{xcolor}
\usepackage{bbding}
\usepackage{enumitem}
\usepackage{pifont}

\usepackage{hyperref}
\usepackage[hyphenbreaks]{breakurl}

\usepackage[misc]{ifsym}
\newcommand{\corr}{(\Letter)}
\usepackage{mwe}
\begin{document}

\title{GeoViz: A Multi-View Visualization Platform for Spatio-temporal Knowledge Graph}
\titlerunning{GeoViz}

\author{Jianping Zhou\inst{1} \and
Junhao Li\inst{1} \and Guanjie Zheng\inst{1}\corr \and Yunqiang Zhu\inst{2}\and 
Xinbing Wang\inst{1}
\and Chenghu Zhou\inst{2}}

\authorrunning{Jianping Zhou et al.}
% If there are two authors, write 'A.L. Benjamin and C.C. Broadus Jr.'
% If there are more than two authors, '[...] et al.' is used.

\institute{Shanghai Jiao Tong University, Shanghai 200240, China 
% \email{\{jianpingzhou,Lijunhao\_hz,gjzheng,xwang8\}@sjtu.edu.cn}
\and
Chinese Academy of Sciences, Beijing 100101, China 
% \email{zhuyq@igsnrr.ac.cn},\email{zhouch@lreis.ac.cn}
\email{\{jianpingzhou,gjzheng\}@sjtu.edu.cn}
}

\maketitle              % typeset the header of the contribution

\begin{abstract}
In this paper, we propose a multi-view visualization technology for spatio-temporal knowledge graph(STKG), which utilizes three distinct perspectives: \textit{knowledge tree}, \textit{knowledge net}, and \textit{knowledge map}, to facilitate a comprehensive analysis of the STKG.
The knowledge tree enables the visualization of hierarchical interrelation within the STKG, while the knowledge net elucidates semantic relationships among knowledge entities. 
Additionally, the knowledge map displays spatial and temporal distributions via spatial maps and time axes, respectively.
Our visualization technology addresses the limitations inherent in single-view approaches and the deficiency of interaction in spatio-temporal perspectives evident in existing visualization methods.
Moreover, we have encapsulated this technology within an integrated, open-source platform named \textsf{GeoViz}.
A demo video of \textsf{GeoViz} can be accessed at \url{https://github.com/JeremyChou28/GeoViz}.
\keywords{Spatio-Temporal Knowledge Graph  \and Multi-View Visualization.}
\end{abstract}

\section{Introduction}
With the exponential expansion of spatio-temporal data, research endeavors have increasingly turned to spatio-temporal knowledge graphs (STKG) for efficient knowledge management and downstream knowledge services. 
However, there is always a lack of effective visualization tools for STKG, which hampers intuitive comprehension of the relationship, distribution, and evolutionary rules within spatio-temporal knowledge triplets.

\textbf{Related Work.} 
Recently, there are many visualization tools available to assist in understanding data and knowledge graphs, such as D3.js, Vis.js, Echarts, and Antv G6~\cite{bostock2011d3,li2018echarts}.
D3.js visualizes data through SVG, canvas, and HTML. 
Vis.js provides a variety of network and timeline chart types.
Echarts facilitates data analysis through numerous diagram types.
% through a large number of diverse diagram types.
Antv G6 specializes in graph visualization.
% To sum up, it is unfortunate that these tools are unable to achieve customized visualization demands in STKG.
Unfortunately, these tools do not support user interaction from the spatio-temporal dimension.
Neo4j~\cite{webber2012programmatic} provides a front-end to visualize its stored knowledge graph, but its visualization is single-view, falling short in displaying sufficient information on the unique spatial and temporal dimensions. 
Moreover, it needs to interact through \textit{Cypher} programming language, which is inconvenient for users to utilize.
% which is unfriendly to users.

\textbf{Our Approach.}
To address the aforementioned issues, we propose a multi-view visualization method to enhance the analysis of STKG. 
Specifically, we design three visualization modes tailored for STKG: \textit{knowledge tree}, \textit{knowledge net}, and \textit{knowledge map}. 
The knowledge tree delineates the hierarchical structure of STKG, the knowledge net elucidates the semantic relationships, and the knowledge map displays the spatial and temporal distributions.
Furthermore, we have developed an integrated platform, \textsf{GeoViz}, which supports the customization of these visualization modes according to user-specific needs. 
For the convenience of users, we devise an easy-to-use web user interface with no-code interaction.

\section{System Overview}
In this section, we provide an overview of three distinct types of visual analysis: \textit{knowledge tree}, \textit{knowledge net}, and \textit{knowledge map}.
Detailed descriptions of these visualization modes and their functionalities are elaborated in subsequent parts.

\subsection{Knowledge Tree}
To elucidate the hierarchical structure of STKG, we design a knowledge tree visualization mode to display the hierarchical interrelations among spatio-temporal knowledge entities.
\textsf{GeoViz} maps the spatial and temporal attributes of the knowledge entities, after normalizing them, to the corresponding nodes of the spatial and temporal concept tree that match their own attributes.
In particular, we define the temporal attribute of the knowledge entity at the root node as the entire timeline, with a first-level sub-nodes as a decade and a secondary sub-nodes as a specific year.
We define the spatial attribute of the knowledge entity at the root node as the world, with a first-level sub-nodes as a continent, and a secondary sub-nodes as a specific country.
When knowledge entities are mapped to each node of the concept tree individually, \textsf{GeoViz} will display a tree-like structure, revealing the hierarchical relationships between knowledge entities.
\subsection{Knowledge Net}
Due to the complex and diverse semantic relationships of STKG, we design the knowledge net visualization to present different correlations.
Furthermore, in order to visualize the region of user's interest from the knowledge net with massive nodes or edges, we can extract the subgraphs of the interested nodes.
In addition, we compute the semantic similarity between the knowledge entities with the assistance of a pre-trained large language model API to discover whether there is any correlation relationship between the unassociated nodes.

\begin{figure}[!t]
    \centering
    \includegraphics[width=\linewidth]{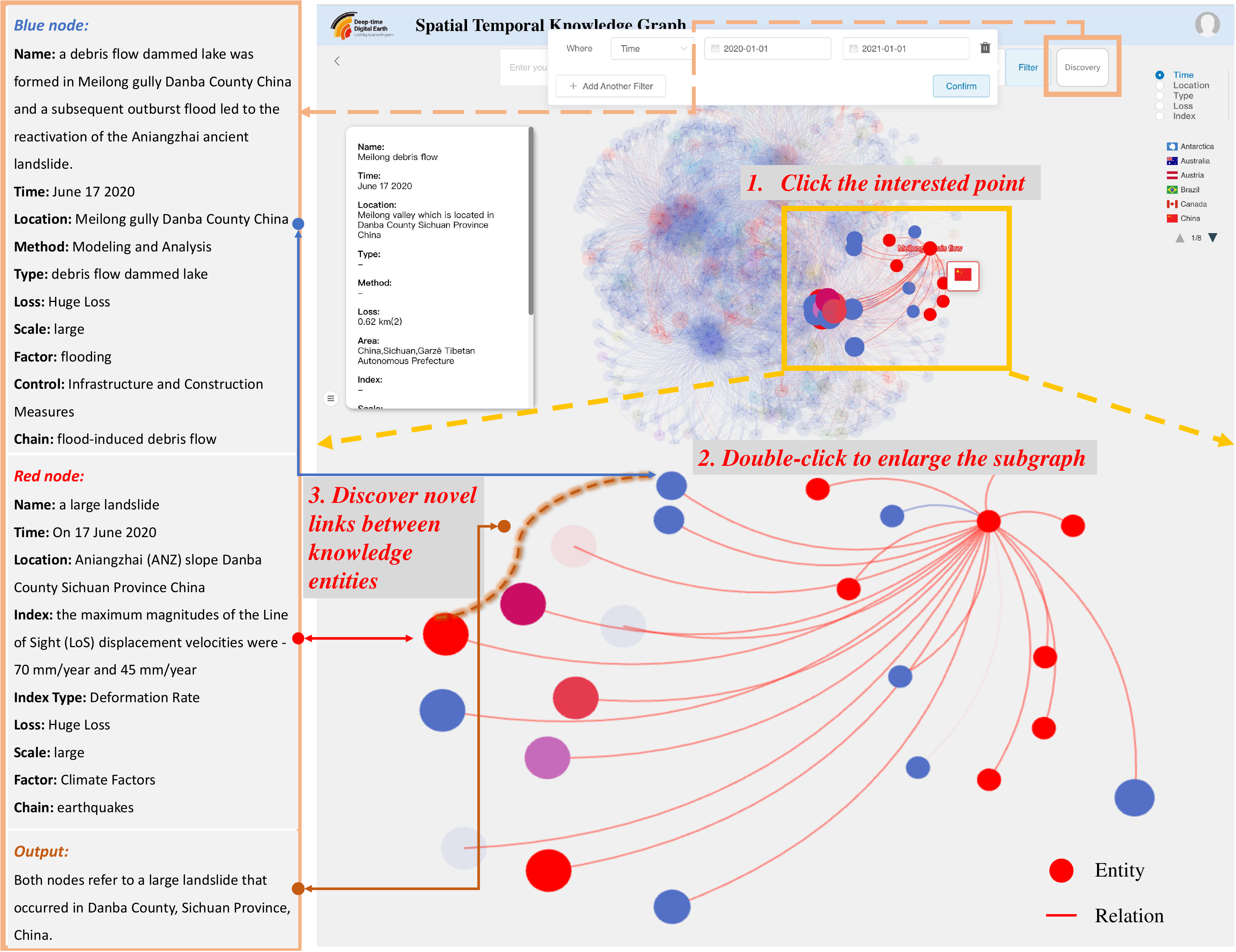}
    \caption{
    An example of knowledge net visualization within the STKG of mountain hazards.
    % Using knowledge net to visualize the spatio-temporal knowledge graph of debris flow hazards and discovering the novel links between triplets.
    }
    \label{fig:knowledge_net}
    % \vspace{-1em}
\end{figure}

We present an example of knowledge net visualization within the STKG of mountain hazards, as shown in Figure~\ref{fig:knowledge_net}.
Specifically, users choose the time range through the filter, and then find the knowledge entities interested through interactive operations. 
Subsequently, users can double-click to zoom-in on the interest node and its subgraph. 
On this basis, users can click the \textit{discovery} button to display new knowledge links (\textit{dashed lines: Both nodes refer to a large landslide that occurred in Danba County, Sichuan Province, China.
}) between the automatically discovered blue node (\textit{a debris flow disaster instance}) and red node (\textit{a landslide instance}).

\subsection{Knowledge Map}

To reveal the spatial and temporal distribution patterns of knowledge entities within the STKG, our system introduces a knowledge map comprising both a spatial map and temporal axes. 
Knowledge entities are systematically positioned within the knowledge map according to their respective time scales and geographic coordinates. 
This arrangement facilitates clear observation of the spatial and temporal distribution patterns, enhancing user comprehension.

\begin{figure}[!t]
    \centering
    \includegraphics[width=\linewidth]{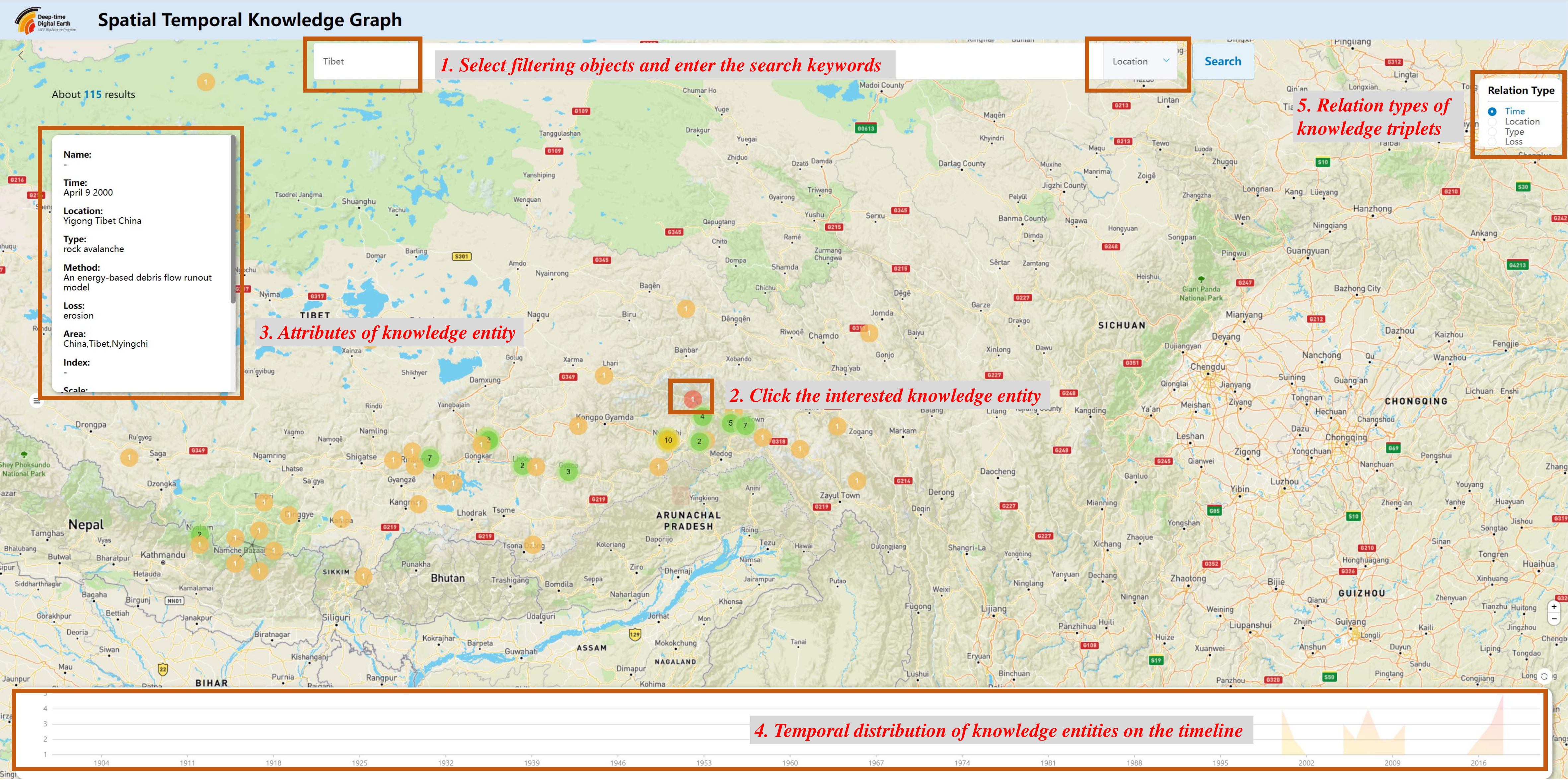}
    \caption{An example of knowledge map visualization pertaining to STKG of mountain hazards.
    % Using knowledge map to visualize the spatio-temporal knowledge graph of debris flow hazards.
    }
    \label{fig:knowledgemap}
    % \vspace{-1em}
\end{figure}

As shown in Figure~\ref{fig:knowledgemap}, we present an example of knowledge map visualization pertaining to the STKG of mountain hazards.
Firstly, users can retrieve knowledge entities by applying filtering criteria and searching keywords, subsequently accessing the attributes of interested knowledge entity. 
In addition to visualizing the spatial distribution of knowledge entities on the map, the timeline facilitates observation of their temporal distribution.
% Using the spatial map and time axes, we can clearly understand the spatial distribution of knowledge triplets, and the temporal evolutionary rules, respectively.
Our knowledge map visualization is a more intuitive way for scientific research to discover the rules of spatio-temporal distribution.

\section{Conclusion}
In this paper, we propose an integrated multi-view visualization platform, \textsf{GeoViz}, aimed at facilitating comprehensive and clear analysis of spatio-temporal knowledge graph(STKG).
In future work, we plan to expand \textsf{GeoViz} to accommodate super-large scale STKG and to conduct more targeted applications in spatio-temporal fields, e.g. mountain hazards, smart-city, etc., enhancing the role of visualization in scientific research.

\bibliographystyle{splncs04}
\bibliography{reference.bib}
%% Note that this preceding line implies that you store your BibTeX references in a file called 'mybibliography.bib'. If you instead store your references in a file with a different name, for instance 'references.bib', the preceding line should read '\bibliography{references}'. Whatever you do, DO NOT put the file name extension .bib inside the \bibliography command; this will trip up LaTeX compilers. 
%
% If you do not want to use BibTeX, you can also type up the bibliography exactly as you see fit, using the following structure:
\end{document}